\documentclass{eas}
\usepackage{graphicx}
\usepackage{amssymb,amsmath,mathrsfs}
\newcommand{\latin}[1]{\textit{#1}}
  \newcommand{\eg}{\latin{e.g.}}
\newcommand{\project}[1]{\textsl{#1}}
  \newcommand{\gaia}{\project{Gaia}}
  
\newcommand{\tv}[1]{\boldsymbol{#1}}
  \newcommand{\Dvec}{\tv{D}}
  \newcommand{\yvec}{\tv{y}}
  \newcommand{\Yvec}{\tv{Y}}
  \newcommand{\thetavec}{\tv{\theta}}
  \newcommand{\Cinv}{\tv{C}^{-1}}
\newcommand{\transpose}[1]{{#1}^{\mathsf{T}}}
\newcommand{\RA}{\mathrm{RA}}
\newcommand{\Dec}{\mathrm{Dec}}
\newcommand{\like}{\mathscr{L}}
\newcommand{\dd}{\mathrm{d}}
\newcommand{\documentname}{article}

\TitreGlobal{\textsl{Gaia}:~At the frontiers of astrometry}
\runningtitle{Hogg \& Lang:~Telescopes don't make catalogues!}
\begin{document}
\title{Telescopes don't make catalogues!}
\author{David W. Hogg}\address{Center for Cosmology and Particle Physics, New York University \emph{and} Max-Planck-Institut f\"ur Astronomie, Heidelberg}
\author{Dustin Lang}\address{Princeton University Observatory}
\begin{abstract}
Astronomical instruments make intensity measurements; any precise
astronomical experiment ought to involve modeling those measurements.
People make catalogues, but because a catalogue requires hard
decisions about calibration and detection, no catalogue can contain
all of the information in the raw pixels relevant to most scientific
investigations.  Here we advocate making catalogue-like data outputs
that permit investigators to test hypotheses with almost the power of
the original image pixels.  The key is to provide users with approximations
to likelihood tests against the raw image pixels.  We advocate three
options, in order of increasing difficulty: The first is to
\emph{define} catalogue entries and associated uncertainties such that
the catalogue contains the parameters of an approximate description of
the image-level likelihood function.  The second is to produce a
$K$-catalogue \emph{sampling} in ``catalogue space'' that samples a
posterior probability distribution of catalogues given the data.  The
third is to expose a web service or equivalent that can
\emph{re-compute on demand} the full image-level likelihood for any
user-supplied catalogue.
\end{abstract}
\maketitle

In probabilistic inference, the goal is to transform information
gathered by the data-taking device into information about the model or
parameters of interest, with as little loss as possible.  The most
precise methods for inference involve ``forward modeling'' of the
measurements: A model that can generate the data accurately is a good
model, and is constrained by every data element that it
(interestingly) generates.  For this reason, everyone who wants to
perform a precise experiment with a telescope wants to model the image
pixels.  Recently, this has been realised in a range of astrophysics
domains, from weak lensing (\cite{bernstein}) to astrometry
(\cite{anderson}; \cite{lang09fm}).

Unfortunately, since its earliest days, astronomy has been done
through \emph{catalogues}.  Historical data sets were made by eye, but
enormous catalogues such as \project{USNO-B} (\cite{usnob}),
\project{2MASS} (\cite{twomass}), and \project{SDSS} \linebreak[4]
(\cite{sdss}) are based on digital intensity measurements.  Despite
this---for practical purposes---the primary data products of those
surveys are catalogues; the survey teams considered it unimportant to
give easy world-wide access to all the raw imaging pixels.

If an investigator wants to know the $J$-band fluxes of a few million
\project{SDSS} sources using \project{2MASS} (something we often want
to know), the only easy option is \emph{catalogue matching} in which
the investigator makes up heuristic matching conditions, tests them on
a subset of sources, and then runs them on the full catalogues.  For
sources that don't get matched, what has the investigator learned?
Very little, because it is almost impossible without access to the
imaging pixels to determine the sensitivity of the \project{2MASS}
data to the source at that source position, and even if the
sensitivity is nominal, a two-sigma measurement (or even zero-sigma)
is much more constraining than the uninformative statement that the
source did not satisfy the \project{2MASS} catalogue-inclusion
criteria.  Furthermore, even when sources \emph{do} match there are
fundamental issues about what can be inferred (\cite{budavari};
\cite{toe}).

Most astronomical projects can be phrased as hypothesis tests.
Thinking forward to \gaia, one important set of projects will involve
finding streams of stars---linear features in phase space---that are
likely remnants of disrupted stellar clusters or galaxies (\eg, Helmi
this volume).  The detection of these lines of stars can be cast as a
hypothesis test: If we put these stars onto a family of similar
orbits, does our explanation of the data improve?  Another set of
projects will involve refinement of Milky Way parameters; these
involve optimisation of some kind of likelihood, possibly
marginalising out the details of the distribution functions (\eg,
\cite{bovy10ss}).  Another set will involve testing the physical
properties of the velocity-space structure in the Milky Way disk
(\cite{dehnen}; \cite{desimone}; \cite{bovy10mg}); these make
different predictions for the \gaia\ data.  The question of which is
best is---in part---a question about their relative likelihoods under
the data.

This \documentname\ is about astronomical imaging in general and the
data products derived therefrom, but the specific examples will be
\gaia-related.

\section{A definition of uncertainty}

The simplest option for transmitting image information to the
catalogue is to build the catalogue such that hypothesis tests against
it are identical to---or as close as possible to---hypothesis tests
against the raw imaging.  Not only is this possible, it is practical
in some cases.

For example, the current plan for the spectroscopic output of 
\project{SDSS-III} \project{BOSS} is that each individual
\project{BOSS} spectrum will be a list of wavelengths $\lambda_i$ and
at each of those wavelengths a flux value $f_i$ and associated inverse
uncertainty variance $1/\sigma_i^2$.  These fluxes and inverse
variances will not be simply ``best-fit'' values; they will be the
parameters of a model of the likelihood function (\cite{bolton10}):
Imagine that for this spectrum an investigator has two possible models
$m_1(\lambda)$ and $m_2(\lambda)$.  The fluxes $f_i$ and inverse
variances $1/\sigma_i^2$ are constructed such that the natural
$\chi^2$ difference
\begin{equation}
    \Delta\chi^2\equiv
     \sum_i \left[\frac{m_2(\lambda_i)-f_i}{\sigma_i^2}\right]
    -\sum_i \left[\frac{m_1(\lambda_i)-f_i}{\sigma_i^2}\right]
\end{equation}
is as close as possible to the $\chi^2$ difference the investigator
\emph{would have} obtained had he or she performed the $\chi^2$ test
in the raw spectrograph image pixels.  Since $\chi^2$ is related to
likelihood by
\begin{equation}
\Delta\chi^2 \approx -2\,\ln\left[\frac{\like_2}{\like_1}\right]
\quad ,
\end{equation}
this makes the $f_i$ and $1/\sigma_i^2$ the parameters of an
approximation to the pixel-level likelihood function.  It is an
approximation because it assumes Gaussian uncertainty, and small
wavelength-to-wavelength covariance (this latter point is addressed in
detail by \cite{bolton10}).

This suggests a general \emph{definition} for catalogue elements and
their associated uncertainties, which we explicitly advocate here.  In
the imagined case of \gaia, the definition would be this: For each
star on the sky, there will be six (or more) parameters, say
\begin{equation}
\transpose{\yvec}=[\RA,\Dec,\varpi,\mu_{\alpha},\mu_{\delta},v_r]
\end{equation}
and an associated six-by-six inverse covariance matrix $\Cinv$.
Imagine that for this star an investigator has two proposals $\Yvec_1$
and $\Yvec_2$ about the six parameters.  We recommend that the catalogue
entries $\yvec$ and $\Cinv$ be \emph{defined} such that the natural
$\chi^2$ difference
\begin{equation}
    \Delta\chi^2\equiv
     \transpose{\left[\Yvec_2-\yvec\right]}\cdot\Cinv\cdot\left[\Yvec_2-\yvec\right]
    -\transpose{\left[\Yvec_1-\yvec\right]}\cdot\Cinv\cdot\left[\Yvec_1-\yvec\right]
\end{equation}
is \emph{as close as possible} to the logarithmic likelihood ratio
$-2\,\ln[\like_2/\like_1]$, now \emph{marginalised over all
calibration and instrument parameters}, that you would have measured
if you had access to the raw image pixels and racks of metal.  The
proposal is to adopt this definition; adoption of this is essentially
equivalent to obtaining the catalog entries and uncertainties from
Gaussian fits to the marginalized likelihood.

The nuisance-parameter marginalization takes this beyond the
\project{BOSS} plan and makes it challenging.  Marginalisation of a
likelihood requires---in the mathematical sense---a prior probability
distribution function (PDF) over the parameters to be marginalised out
and it looks something like this:
\begin{eqnarray}\displaystyle
\like(\Yvec) & \equiv & p(\Dvec|\Yvec) \nonumber\\
  & = & \int \dd\thetavec\,p(\thetavec)\,p(\Dvec|\Yvec,\thetavec)
\quad ,
\end{eqnarray}
where the object $\Dvec$ is the entire raw data set (pixel
intensities), $\Yvec$ is the set of six parameters of the individual
object, and $\thetavec$ is the set of all calibration parameters in
the model, including all those related to attitude and hardware.  The
unmarginalised likelihood---the probability of the data given the star
and calibration parameters---depends, of course, on both the star and
the calibration parameters; the marginalisation is the only
conservative and accurate way to propagate uncertainties about
calibration into the astrophysical parameters of interest.  The prior
$p(\thetavec)$ is required as a measure on the parameter space that
permits integration.

If the entries are defined this way, a catalogue user can make
approximations to likelihood comparisons at the pixel level.
Unfortunately, this proposal is only sensible when the marginalized
likelihoods are close to Gaussian in form and when star--star
covariances can be ignored.  These covariances are not expected to be
negligible for \gaia\ (see Holl this volume).  However, this will be a
problem for almost all of the currently conceived plans for the \gaia\
catalogue.  Transmission of star--star covariances is permitted by our
next proposal.

\section{A sampling in catalogue space}

At the present day, in most situations of fitting large
(non-parametric, or highly parameterised) models to large data sets,
investigators report uncertainty information through \emph{samplings}.
A sampling has the disadvantage that it is a
mixture-of-delta-functions approximation, but it has the great
advantage that it can, in principle, describe an arbitrarly
complicated PDF.  There are also excellent sampling technologies
available (\eg, \cite{mackay}).

A useful sampling of \gaia\ catalogues could have the following
properties: There would be $K$ catalogues that represent a
sampling---perhaps not a Poisson sampling, but a sampling---from the
posterior PDF in what you might call ``catalogue space''.  Any
experiment or measurement is performed on all $K$ samples.  The
variance across the $K$ outcomes of the $K$ identical experiments is
an estimate of the uncertainty on the results of the outcome on the
primary catalogue.  In this vision, the the sampling provides a
rank-$K$ approximation to the full-catalogue covariance matrix (which
is billions by billions and non-sparse).

Importantly, in this vision, the sampling of catalogues should be a
sampling not just in the space of the astrophysical parameters but
also in the space of the attitude and calibration parameters (the
nuisance parameters).  This ensures that the variance over samples
contains the propagated uncertainty from the nuisance parameters, and
it doesn't add significantly to any of the technical challenges of
producing the sampling.

Although there is an enormous literature on sampling methods for large
problems, there is a useful hack that might be appropriate if
practical sampling methods fail on a problem of this scale:
Approximate samples can be made by inflating leave-one-out jackknife
trials.  The idea would be to cut the raw data stream into $K$ equally
informative (roughly speaking, equal-sized) disjoint subsamples, and
construct the $K$ complementary leave-one-out subsamples, in each of
which one of the $K$ disjoint subsamples is left out.  All data
analysis (from raw data to inference of all nuisance and astrophysical
parameters through to catalogue generation) is performed on each of
the $K$ leave-one-out subsamples.  The \emph{difference} (in catalogue
space) between the primary catalogue and each of the $K$ leave-one-out
catalogues can be amplified by a factor of (very close to) $K$ to
produce a sampling that has the same full-catalogue variance and
covariance as the classical jackknife estimate.  This might be an
expedient way to produce and publish the jackknife estimate of the
full error covariance, that has most of the good properties of a
sampling.

Technically, a sampling is a description not of the likelihood but of
a posterior PDF, and it has the disadvantage to its users that it has
had a prior PDF multiplied in.  This is a fundamental limitation of
the approach, but for most kinds of catalogue outputs astronomers are
interested in, the data are so informative that the posterior PDF in
catalogue space looks very much like the likelihood function unless
the adopted priors are extremely informative.  Another issue is that
additional information (say, from another telescope or instrument),
which is expressed as a multiplicative likelihood term, can cause huge
changes in the relative posterior probabilities of the $K$ catalogue
samples.  Indeed, all $K$ samples may be essentially ruled out by new
information, or the samples may fail to capture an important mode in
catalogue space.  This is a fundamental limitation of sample-based
representations.

In an ideal but challenging world, a sampling in catalogue space would
explore regions of differing \emph{complexity}.  Not sure if a star is
binary?  Some of the $K$ catalogues would have it binary and some
would have it single.  There is no reason in principle that the
catalogues would be qualitatively similar in cases in which there is
real statistical scientific uncertainty.  This complexifies
substantially sampling strategies, and complexifies downstream
analyses by users, so we leave it here only as a comment.  Changes to
catalogue complexity are handled very naturally by our third proposal.

\section{Exposing the full likelihood function}

All this has been about approximations to the likelihood function,
which begs the question: Why not just publish the likelihood function
itself?  Here what we imagine is an interface (perhaps in the cloud;
see O'Mullane this volume) to which a user could submit a
\emph{catalogue diff}---a difference between the primary
\gaia\ catalogue and the catalogue he or she wants to test.  The
machinery behind the interface would use the modified catalogue to
generate the raw pixels, compare to the data, and return the
\emph{likelihood diff}.  Of course the user must be permitted to
modify also nuisance parameters, or else be presented with the option
of obtaining the results marginalised over these.

An interface of this kind would be expensive to build, run, and
maintain; it is probably impractical at the present day.  However, it
would permit literally arbitrary hypothesis testing, and arbitrary
calculation of covariances.  It would also make it the responsibility
of users to perform their own uncertainty analysis and propagation,
relieving the teams of some of their most burdensome duties.
Extremely clever systems could cache computation so subsequent users
could benefit from the users that have come before.  However, without
technology development, this proposal is probably impractical for the
\gaia\ data.

\section{Discussion}

\emph{How big do you have to make $K$?}  There is no simple answer;
this depends on the complexity of the posterior PDF and the precision
with which users need to know their uncertainties.  Our attitude is
that you never---in any finite experiment---know your
\emph{uncertainties} to high precision, so we have the intuition that
$K$ does not have to be large to be an enormous improvement over a
single catalogue with reported single-star uncertainties.  One
order-of-magnitude option for $K$ in the case of \gaia\ is the number
of \emph{visits}, the number of times the spacecraft scans across a
typical star.  Jackknifing more finely than this will not help much on
individual-star uncertainties.

\emph{But if anyone can make their own catalogue, the data will be
  changing; the science won't be repeatable!}  This point is good;
published science needs to be repeatable by referees and subsequent
investigators.  On the other hand, our \emph{understanding} of the
\gaia\ data will necessarily evolve as calibrations and physical
models evolve.  There is no hope for a completely stable catalogue.
What \emph{is} fixed is the set of raw pixel data.  The approaches
advocated here make the raw data the primary objects of release, and
therefore tie the release to the only part of the data analysis that
\emph{is} stable and repeatable.  Of course, it is necessary for
reasons of repeatability to clearly tag, cut, and release
well-documented and stable versions of the data, likelihood function,
and primary catalogue.

\emph{Is there any relevant difference between a Bayesian and a
  frequentist?}  Bayesianism is required for marginalization, because
the prior provides the measure for integration.  However, Bayesians
and frequentists agree---or should agree---that the most important
publication of an experiment is the \emph{likelihood function}.  It is
through the likelihood that data impact our beliefs; different
investigators have different priors, different objectives, and
different external data at their disposal.  They can only combine
these with the \gaia\ data properly if the likelihood function is
exposed.  Even if the \gaia\ outputs end up being a sampling of some
posterior PDF, those samples should come with evaluations of the
prior, so that subsequent users can divide it out before combining
with their own individual knowledge and data.

\acknowledgements It is a pleasure to thank Anthony Brown, Francoise
Crifo, Ken Freeman, and Lennart Lindegren, whose comments
\emph{during} the meeting shaped these ideas in real time, and Coryn
Bailer-Jones, Jo Bovy, and Sam Roweis for valuable discussions.
Partial financial support was provided by NASA (grant NNX08AJ48G), the
NSF (grant AST-0908357), and a Research Fellowship of the Humboldt
Foundation.

\end{document}